\documentclass[apjl]{emulateapj}

\lefthead{L. Ben-Jaffel}
\righthead{HD209458b: inflated but not evaporating}

\def \lya\ {Lyman-$\alpha\, $} 

\begin{document}

\title{Exoplanet HD209458b: inflated hydrogen atmosphere \\ but no sign
 of evaporation}

\author{Lotfi Ben-Jaffel} 
\affil{Institut Astrophysique de Paris, UPMC, CNRS, 98 bis Blvd Arago,
 75014 Paris, France}
\email{bjaffel@iap.fr}

\slugcomment{Astrophysical J. Letters, v. 671, pp 61-64, 2007}
\received{2007 August 21}
\accepted{2007 October 18}
\published{2007 November 8}
\begin{abstract}

Many extrasolar planets orbit closely to their parent star. Their existence raises the fundamental problem of loss and gain in their mass. For exoplanet HD209458b, reports on an unusually extended hydrogen corona and a hot layer in the lower atmosphere seem to support the scenario of atmospheric inflation by the strong stellar irradiation. However, difficulties in reconciling evaporation models with observations call for a reassessment of the problem.  Here, we use HST archive data to report a new absorption rate of $\sim 8.9\%\pm 2.1$\% by atomic hydrogen during the HD209458b transit, and show that no sign of evaporation could be detected for the exoplanet. We also report evidence of time variability in the HD209458 \lya\ flux, a variability that was not accounted for in previous studies, which corrupted their diagnostics. Mass loss rates thus far proposed in the literature in the range $5\times 10^{10}-10^{11}\, {\rm g\, s^{-1}}$ must induce a spectral signature in the Lyman-$\alpha$ line profile of HD209458 that cannot be found in the present analysis. Either an unknown compensation effect is hiding the expected spectral feature or else the mass loss rate of neutrals from HD209458 is modest.

\end{abstract}


\section{Introduction}
Of all the planets discovered outside the realm of our solar system, some of the most dramatic new classes of objects are those in which the planet is a gas giant orbiting at merely a few stellar radii ($\sim 0.02 {\rm AU}$) from its parent star. These close-in extrasolar planets are Jupiter-like giants that are exposed to strong fluxes, magnetic fields, and plasma winds---a very harsh and active stellar environment. Because of their stars' proximity, gravity, through tidal effects, distorts the shape of their atmosphere while the continuous extreme ultraviolet (UV) energy deposition inflates it (\cite{lam03}; \cite{bar04}; \cite{lec04}; \cite{yel04}; \cite{jar05}; \cite{tia05}; \cite{mun07}; \cite{vil07}). Unfortunately, little is known about those regions that separate extrasolar giant planets from their stars, particularly the immediate environment of the planet. 

One of the most extensively studied extrasolar systems is HD209458. For
reference, HD209458b was first discovered transiting its parent star and
covering $1.5$\% of its disk (\cite{cha00}; \cite{hen00}).  Some of the
first attempts to learn about the immediate environment of the planet
were \lya\  ($121.6$ nm) observations of the system using the Space
Telescope Imaging Spectrometer (STIS) onboard the Hubble Space Telescope
(HST).  A first program of observation, obtained with the STIS/G140M
grating and the 52"x0.1" slit, was implemented in $2001$ during HD209458
planetary transit but no conclusions were reported (see
Table~\ref{tbl-1}). Soon after, a second program visited the target
during three transits (\cite{vid03}).  An initial analysis of this data
set concluded that a huge cloud of hydrogen is covering $15\%\pm 4$\% of
the stellar disk (\cite{vid03}); it also claimed that spectral
absorption during transit is deeper on the blue side of the stellar
line.  Accordingly, the hydrogen cloud was required to extend beyond the
planetary Roche limit where an intense escape of $\sim~ 10^{10}\, {\rm
g\, s^{-1}}$ of hydrogen is a priori operating. These results, along
with other far UV low-resolution observations of heavy constituents, led
to the conclusion that the upper atmosphere of HD209458b should be in a
hydrodynamic blow-off state (\cite{vid03, vid04}). 

Numerous studies then followed on different mechanisms for hydrogen loss
from hot exoplanets closely orbiting their stars (\cite{lam03};
\cite{bar04}; \cite{jar05}; \cite{lec04}; \cite{yel04, yel06}; \cite{tia05}; \cite{mun07}). 
As noted by \cite{mun07}, all loss rates thus far proposed by theoretical models in the range $5\times 10^{10}-10^{11}\, {\rm g\, s^{-1}} $ exceed the lower limit provided by \cite{vid03}. Unfortunately,  most studies neglected to quantitatively translate their loss rate to a spectral absorption in the \lya\ line profile that could be tested with the HST observation. Independently, pointing out that the observed mass function distribution of extrasolar giant planets (EGPs) follows a trend ${\rm M^{-1}}$ for mass range 
$\sim 0.2-5\, {\rm M_{J}}$, where ${\rm M_{J}}$ is the Jovian mass, \cite{hub07} derived the same mass function distribution for highly irradiated EGPs orbiting at distances smaller than $\sim 0.07$ AU. Accordingly, \cite{hub07} rejected substantial mass loss during EGPs' migration to smaller distances from their star, unless the loss mechanism is compensated by an unknown process. When combined with the unusual scales derived for the hydrogen extent and escape, all these studies then call for a careful reassessment of the HST \lya\ observations thus far obtained on HD209458, at least in order to provide validated constraints on theoretical models.

\section{Observations and data analysis}

In the following, we report a new analysis of archive data obtained
during the two HST/STIS programs described above. In total, we have four
visits of the target corresponding to three exposures of roughly $\sim
2000 \, {\rm s}$ duration each, resulting in $12$ exposures of the systems around the transit period (Table~\ref{tbl-1}). All observations were obtained in the time-tag mode, a technique that keeps track every $125\times 10^{-6}\, {\rm s}$ of photon events during each exposure. The question, then, is: why is this mode of observation important in the present case? First, we stress that the transit effect is a weak variation of the stellar signal. As such, its trend is best represented by a dense time series.  Second, chromospheric and coronal variabilities of the star are unknown in the \lya\ spectral  window considered here, and this may seriously corrupt any diagnostic. 

\begin{table*}
\tablenum{1}
\caption{HST/STIS Data Set on HD209458 Used in This Study. All observations were obtained with the G140M grating and the 52"x0.1" long slit. Transit central time (TCT) is defined by 2,452,826.628521 HJD (\cite{bal06}; \cite{knu07}).  \label{tbl-1}} 
\begin{center}
\begin{tabular}{cccccc}

Dataset name &	Program ID &	Start time - TCT (s) &	Duration (s) &	End
 time - TCT (s) \\
O4ZEA4010 & 7508	& -7980.28  & 2600.	& -5380.12 \\
O4ZEA4020 & 7508	& -2202.29  &	2600.	& 397.87   \\
O6E201010 & 9064	& -8075.11	& 1780.	& -6295.02 \\
O6E201020 & 9064	& -2246.12	& 2100.	& -145.96  \\
O6E201030 & 9064	& 3531.89	  & 2100.	& 5631.98  \\
O6E202010 & 9064	& -10167.67	& 1780.	& -8387.50  \\
O6E202020 & 9064	& -4750.68	& 2100.	& -2650.52  \\
O6E202030 & 9064	& 1025.31	  & 2100.	& 3125.46  \\
O6E203010 & 9064	& -11258.45	& 1780.	& -9478.27  \\
O6E203020 & 9064	& -5876.47	& 2100.	& -3776.29  \\
O6E203030 & 9064	& -102.45	  & 2100.	& 1997.73  \\

\end{tabular}
\end{center}

\end{table*}

To properly handle time tagged data, we partition each time-tag exposure
into a set of shorter sub-exposures, taking into account the
heliocentric and barycentric time correction procedure in which
ephemerides are retrieved from the archives before the IRAF/STSDAS
"odelaytime" procedure is applied (\cite{bro02}). After several trials,
we found that $300 \, {\rm s}$ sampling of the time-tag data is a good
compromise between acceptable signal-to-noise ratio (S/N) and time coverage. Next, each sub-exposure is calibrated through the STIS pipeline.  The emitting source (the star plus sky background) is presumed extended, an option that allows efficient control of the subtraction of the sky background contamination. Time is then converted to fixed orbital phases measured from the transit central time (TCT), itself carefully taken from the most recent and accurate determination of the HD209458 system parameters (\cite{bal06}; \cite{knu07}). Sub-spectra of identical phase positions are accumulated from the initial twelve exposures, resulting in a unique 53 bins time series of the system versus the orbital phase angle (Fig~\ref{fig1}). Unexpectedly, three gaps, lasting respectively $312 \, {\rm s}$, $404 \, {\rm s}$, and $406 \, {\rm s}$, appear in the time series, for which no observation is available. Because the three gaps are narrow and well-separated from each other, we determined that filling them has a negligible effect on our final conclusions (\cite{sch01}). Errors due to photon counting have been propagated, taking into account the correlation between the different phase positions relative to the initial sampling of sub-exposures time over the full observing time period. 

We next define the wavelength domain of contamination by the sky
background, including both the Earth's geocorona and the interplanetary
medium emissions. The difficulty comes from the uncertainty about the
geocorona's strength when estimated from a detector sector, along the
STIS slit, different from the one where the stellar signal was
recorded. First, we subtracted the dark noise of the detector following
(\cite{vid03} and  \cite{bal06}) and then compared the sky background signal from different sectors along the slit and for different conditions of observation. Our conclusion is that the STIS MAMMA detector has an inherent non-uniformity corresponding to an incompressible uncertainty of $5$\% on extended sources. Coincidentally, this uncertainty is comparable to photon statistical errors. To ensure that such error will not corrupt the stellar signal per wavelength pixel at the $1$\% level, we deduce that a void window $[121.541, 121.584]$ nm should be disregarded in any spectral analysis that requires high accuracy, such as for a transit event or short-term stellar variability.

\begin{figure}
\epsscale{1.4}
\includegraphics [scale=0.56]{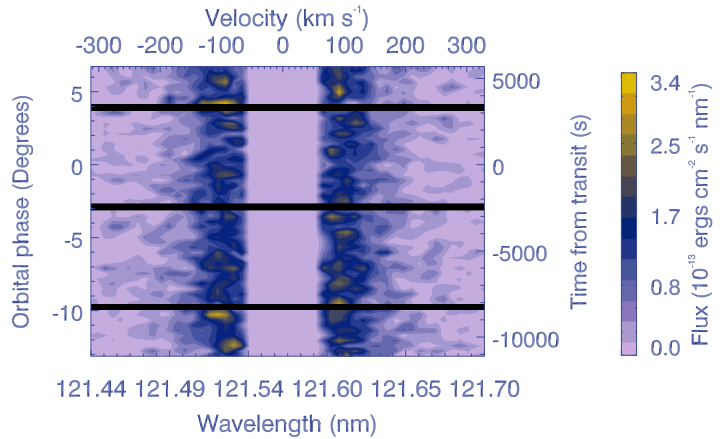}
\caption{Contours of equal flux for the HD209458 time series vs 
 wavelength (nm) and time from transit (or planetary orbital phase
 angle). Top axis shows velocities in the stellar rest frame ($\sim 10\,
 {\rm km\, s^{-1}}$) from heliocentric system). Horizontal dark bands
 are time gaps of $\sim 300$-$400$ ${\rm s}$ for which no observation is
 available. These bands are linearly interpolated using nearby spectra
 for light curves analysis and in/out transit spectra
 comparison. Following the time evolution of the signal (from bottom to top), we observe the transit absorption as a slight dimming of the stellar flux starting $\sim 5000\, {\rm s}$ before TCT.  The wide, vertical, light purple band corresponds to the spectral window of extinction of the stellar signal by the interstellar gas along the line of sight (\cite{woo05}}
\end{figure}

\section{Light curve trend and time variability}

Next, to obtain the trend of the planetary transit in \lya\ , we integrated the time series spectra in the range $[121.483, 121.536]$ nm on the blue wing and $[121.589, 121.643]$ nm on the red wing.  These ranges were selected so that the stellar signal per wavelength pixel remains above the statistical noise ($\geq 1\, \sigma $). The resulting light curve is noisy, but a trend is apparent and can be efficiently extracted (Fig. 2a). To dampen the signal noise while keeping a clear trend, the best compromise is to gather data by eight phase bins for a new bin of $2400\, {\rm s}$. Accumulating the signal from three new bins ($3\times 2400$ s) inside transit, we derive $\sim 8.9\pm 2.1$\% drop-off of the stellar \lya\ intensity during the planetary transit (Fig. 2a). Our absorption rate of $\sim 8.9\%\pm 2.1$\% is much lower and accurate than reported in a previous study (\cite{vid03}), yet a marginal agreement could be found between our maximum rate (11\%) and their bottom value. If this obscuration is converted directly to a planetary occulting disk, then one would obtain a hydrogen cloud of $\sim 2.47\pm 0.30$ R$_{\rm P}$ radius, much smaller than the Roche lobe limit of $\sim 4.08$ R$_{\rm P}$ (\cite{gup03}), where R$_{\rm P}=1.32$ R$_{\rm J}$ is the most recent estimate of the radius of HD209458b (\cite{bal06};  \cite{knu07}), and R$_{\rm J}$ is the Jovian one. Now, to capture the trend of the transit curve, we used a sophisticated 2D model of planetary transit at \lya\ that accurately accounts for the atmospheric radial structure of the planet (\cite{yel04};  \cite{ben07}) and properly estimates the atmospheric obscuration versus wavelength, including extinction by the interstellar gas intervening along the line of sight (\cite{woo05}). Our best least square fit is shown in Figure 2a. For our purpose of time analysis, we remark that a functional fit could also be a good model to obtain the light curve's trend.

\begin{figure}
\epsscale{1.4}
\plotone{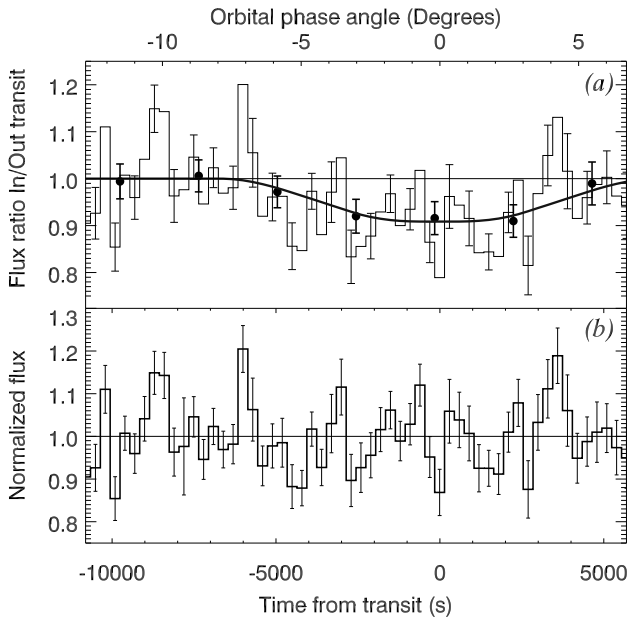}
\caption{{\bf (\it a)} Light curve (LC) obtained from our time series
 (as shown in Fig. 1) by integration of the signal in the spectral
 windows [121.483, 121.536] nm and [121.589, 121.643] nm.  Histogram and
 related errors (plotted every two bins for clarity) show the LC with a time
 bin of $300\, {\rm s}$, while filled circles with attached error bars
 represent the LC rebinned to a larger timescale of $8\times 300 {\rm s}=
 2400 {\rm s}$.  The solid curve is our best least square fit to the
 rebinned LC. A total obscuration of $\sim 8.9\%\pm 2.1$\% is derived
 during the planetary transit. {\bf (\it b)}  Ratio of LC (histogram) to
 best model fit (solid line) is shown with related statistical error
 bars. This ratio cancels the transit trend shown in ({\it a}). The resulting signal is a good indicator of the variability of the HD209458 \lya\ intensity vs time. \label{fig2}}
\end{figure}

We can now determine the stellar signal time evolution after we cancel the transit trend using our best fit to the observed light curve (Fig. 2a).  The resulting ratio shows a variable behavior with an average amplitude $\sim 8.6 \pm 5.6$\% of the stellar integrated intensity (Fig. 2b). Using the Durbin-Watson statistical test (\cite{dur51}), we found no apparent serial correlation at the 1\% confidence level in the corrected signal---a signal that also shows no evident periodicity.  HD209458 was previously suspected to have a relatively moderate chromospheric activity from CaII H and K lines that were recorded over full orbits of the system (\cite{shk05}).  Our finding of a time variation of 8.6\% on average in the stellar \lya\ signal, with peaks that may reach $\sim 20$\% ($> 3\sigma\, $), seems to support a relatively active corona of the star, presumably up to the planet's orbit. Such activity could be of common origin (flaring, non-uniformity of the stellar disc during transit, etc.) and/or related to an enhancement of magnetic activity on the star-planet line (\cite{zar07}). Also, one can speculate about the hydrogen cloud topology around the planet and its evolution with time. To that end, comparative studies with interacting binary stars may be useful in clarifying the different regimes of interaction between an exoplanet and its host star (\cite{sho94}). Unfortunately, the FUV observations thus far obtained do not cover a full orbit of the planet, thereby making it difficult to predict the exact configuration of the star-planet system. In any case, we believe that the unusual $15$\% obscuration previously reported (\cite{vid03}) was corrupted by this unaccounted-for variable component in the star-planet system signal.  Here we can extract it because we are able to sample the transit period by a dense time series using the information gathered from the time tag mode of HST/STIS and $\sim 25$\% more observation time from the archives. 

\section{Planetary mass loss or flux variability?}

In the following, we compare the in/out of transit stellar line profiles. The impetus of this study is the need to determine the relevance of a blueshifted absorption in the stellar line profile that may occur during transit, as claimed in earlier studies (\cite{vid03}). On the one hand, we derive an average unperturbed profile of the HD209458 \lya\ emission line by merging all sub-spectra of the time series that we correct for the transit trend with the best fit shown in Figure 2a. The resulting profile is a good reference that best represents the out-transit stellar line and for which time variability has been reduced to the 1\% signal level (Fig. 3a). On the other hand, the in-transit line profile, when corrected for the $\sim 8.9 $ \% drop-off during transit, properly recovers the unperturbed line profile (Fig. 3a), leaving no real possibility of extra absorption as claimed in prior studies (\cite{vid03}). 

\begin{figure}
\epsscale{1.2}
\plotone{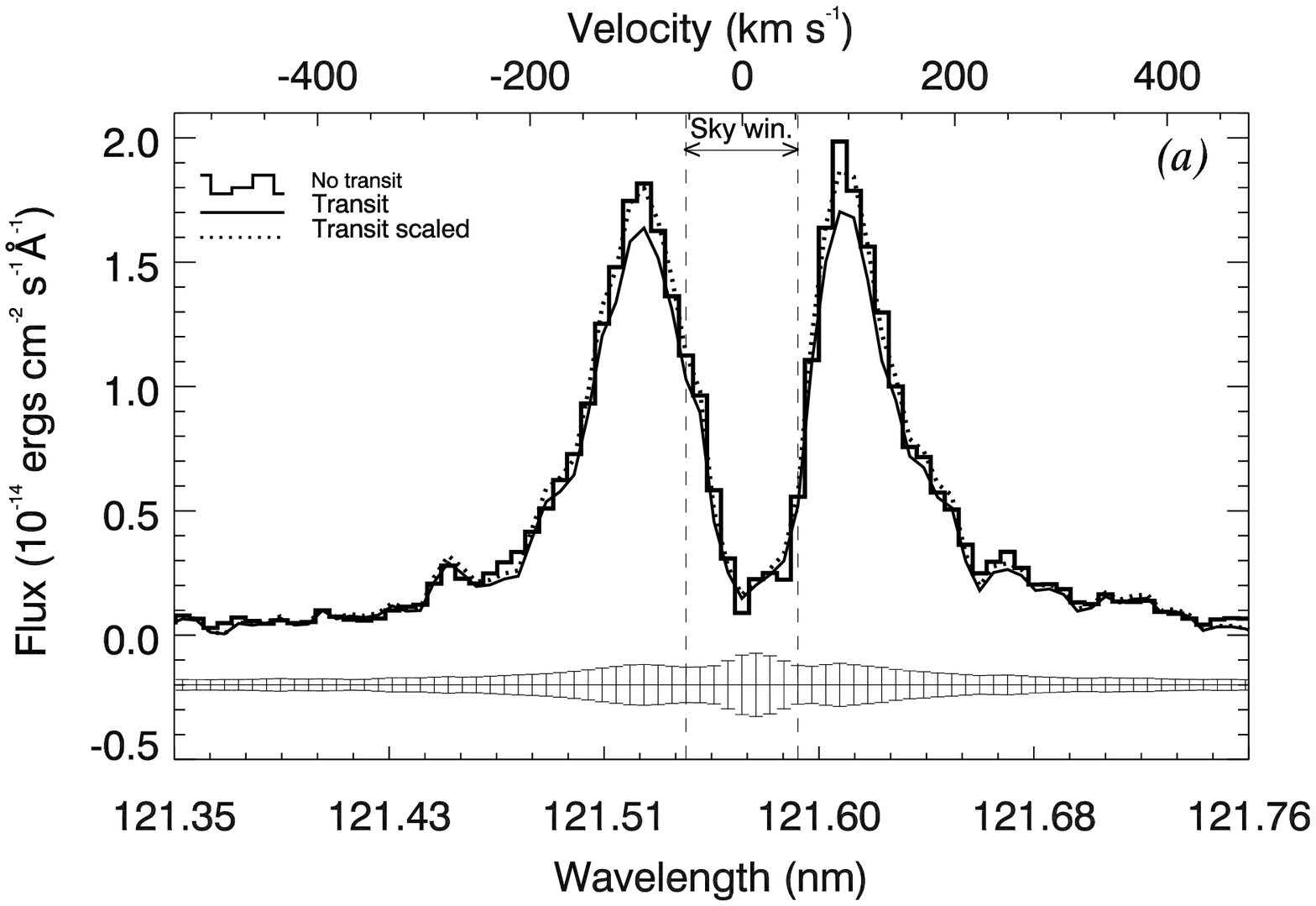}
\plotone{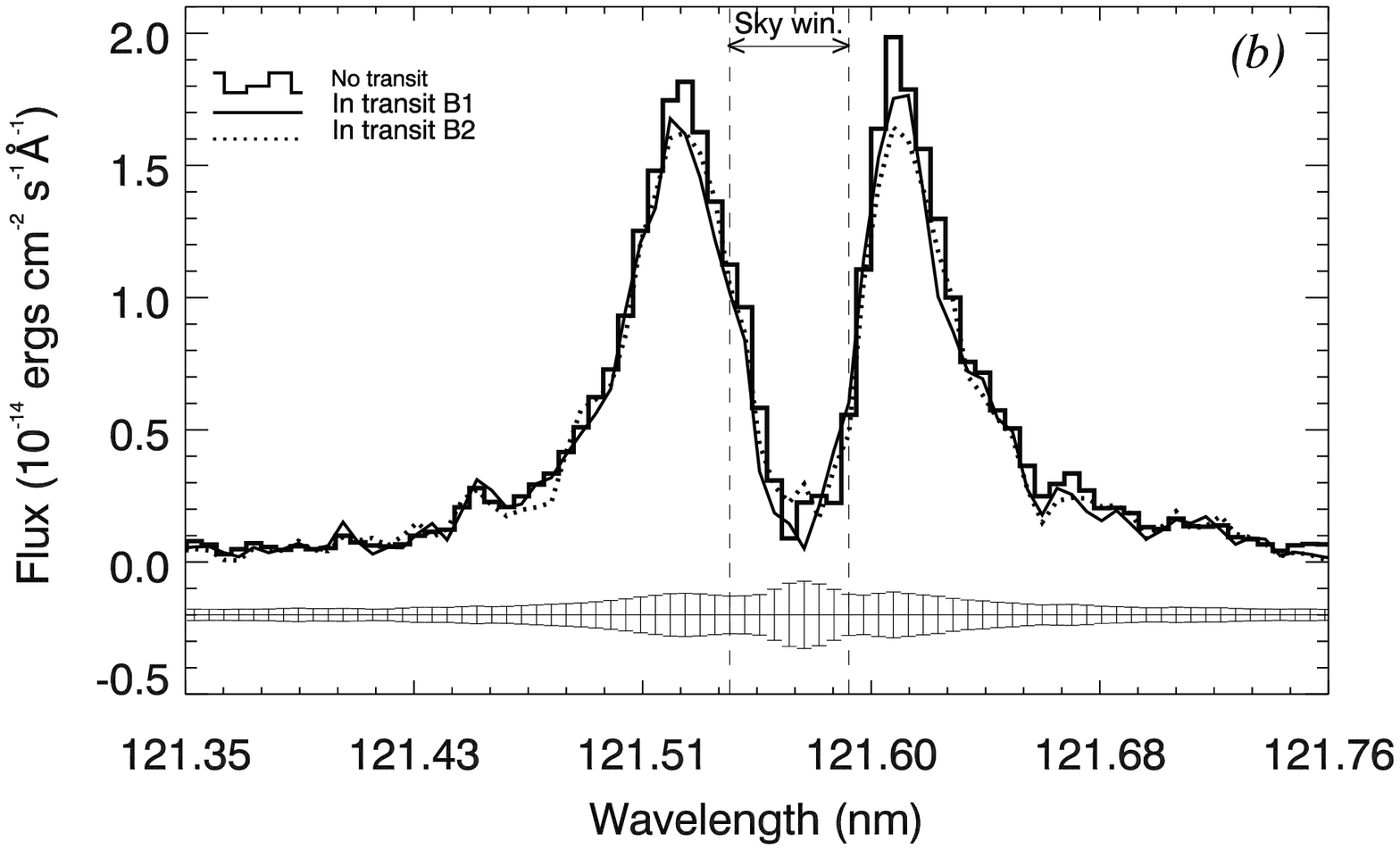}
\caption{Comparison between \lya\ line profiles in and out of transit
 period. The sky background spectral window is indicated by two dashed
 vertical lines. {\bf (\it a)} The in-transit line profile (solid thin line)
 is accumulated for the time period starting $\sim 3900\, {\rm s}$ before TCT
 and ending $\sim 3900\, {\rm s}$ after it. To correct for the $\sim 8.9$ \%
 obscuration derived in this study, the corresponding intensity is
 scaled by $1.098$.  The resulting line profile (dotted curve) properly
 recovers the unperturbed line profile (histogram). {\bf (\it b)} The first
 in-transit line profile, B1 (thin solid line), was accumulated over the
 time period starting $\sim 4000$ s before TCT and
 ending $\sim 600\, {\rm s}$ after it.  The second in-transit line
 profile, B2 (dotted line), was accumulated over the time period
 starting $\sim 1800 {\rm s}$  before TCT and ending $\sim 3900 {\rm s}$ after it.\label{fig3}}
\end{figure}

To further investigate how time variability of the HD209458 \lya\ emission line corrupts the diagnostic as it pertains to extra absorption
or emission features that may appear in the stellar line during transit,
we selected two phase windows inside the transit period for which we
compared line profiles to the unperturbed stellar line. As shown in
Figure 3b, a direct comparison would indicate that line peaks are
equally absorbed for line profile B1, while for line profile B2, the red
peak is the most absorbed. On the basis of line profile B2, the
diagnostic would be just the opposite of that of (\cite{vid03}), leading
to escaping hydrogen toward the star, while for line profile B1, the
diagnostic would be no H escape. The problem is that these
interpretations of preferred blueshifted or redshifted absorption do not
account for the relatively strong modulation of the stellar signal
evidenced in this study. Therefore, any claim of a preferred absorption
during transit, either blue or redshifted, is not realistic,
particularly at this relatively modest level of the signal to noise. It
follows that the blueshifted absorption, advanced in previous studies
(\cite{vid03}; \cite{lec04}) as a signature of atmospheric evaporation
in a cometary-like tail of HD209458b, has, unfortunately, no foundation
in the HST/STIS data set as it was only the effect of the stellar signal variability with time that corrupted the diagnostic. 

\section{ Conclusion} 

We use HST archive observations of the \lya\ emission of HD209458 to report an absorption rate of $\sim 8.9\% \pm 2.1$\% by atomic hydrogen during the transit of the planetary companion. If the planet is sketched as a compact blocking body, our analysis requires an H cloud effective extent that does not exceed $\sim 2.5$ R$_{\rm P}$---a size that falls short of the Roche limit $\sim 4.08$ R$_{\rm P}$  of HD209458b. In addition, time variability of the stellar flux is evidenced, but no sign of extra or Doppler-shifted absorption could be detected during transit. 
This absence of extra absorption during transit and the relatively small size of the effective area of the hydrogen cloud around the exoplanet make it difficult to conceive of significant atmospheric evaporation from the planet. Of course, we cannot rule out that a complex atmospheric distribution, related to a particular planet-star interaction scenario, may hide or compensate the loss signature during the observing time. Future HST (when repaired) FUV observation of the system during a full planetary orbit should help to disentangle the different processes in play.

\acknowledgments
 The author acknowledges support from Université Pierre et Marie Curie (UPMC) and the Centre National de la Recherche Scientifique (CNRS) in France. This work is based on observations with the NASA/ESA Hubble Space Telescope, obtained at the Space Telescope Science Institute, which is operated by AURA, Inc.


\clearpage

\begin{thebibliography}{}

\bibitem[Ballester et al., (2007)]{bal06} Ballester, G., Sing, D., Herbert, F. 2007, Nature, 445, 511.
\bibitem[Baraffe et al., (2004)]{bar04} Baraffe, I., et al. 2004, \aap, 419, L13.
\bibitem[Ben-Jaffel et al., (2007)]{ben07} Ben-Jaffel, L., Kim, Y., and Clarke, J. 2007, Icarus, 190, 504.
\bibitem[Brown et al., (2002)]{bro02} Brown, T., et al. 2002, in HST STIS Data Handbook, B. Mobasher, Ed. (Baltimore, STScI, v 4), pp 131-192.
\bibitem[Charbonneau et al., (2000)]{cha00} Charbonneau, D., et al. 2000, \apj, 529, L45.
\bibitem[Durbin and Watson (1951)]{dur51} Durbin, J., and Watson, G.S. 1951, Biometrika 38, 159.
\bibitem[Gu et al., (2003)]{gup03} Gu, P., Lin, D., and Bodenheimer, P. 2003, \apj, 588, 509.
\bibitem[Henry et al., (2000)]{hen00} Henry, G. et al. 2000, \apj, 529, L41.
\bibitem[Hubbard et al., (2007)]{hub07} Hubbard, W.B., et al. 2007, \apj, 658, L59.
\bibitem[Jaritz et al., (2005)]{jar05} Jaritz, G., et al. 2005, \aap, 439, 771.
\bibitem[Knutson et al., (2007)]{knu07} Knutson, H., et al. 2007, \apj, 655, 564.
\bibitem[Lammer et al., (2003)]{lam03} Lammer, H., et al. 2003, \apj, 598, L121.
\bibitem[Lecavelier et al., (2004)]{lec04} Lecavelier des Etangs, A., et al. 2004, \aap, 418, L1.
\bibitem[Munoz (2007)]{mun07} Munoz, G. A. 2007, Planet. Space Sci., 55, 1426.
\bibitem[Schneider (2001)]{sch01} Schneider, T. 2001, J. Climate, 14, 853.
\bibitem[Shkolnik et al., (2005)]{shk05} Shkolnik, E., et al. 2005, \apj, 622, 1075.
\bibitem[Shore et al., (1994)]{sho94} Shore, S., Livio, M., and van den
				  Heuvel, E. 1994, in Interacting
				  binaries, ed. Nussbaumer, H., and Orr
				  (Berlin: Springer), 1
\bibitem[Tian et al., (2005)]{tia05} Tian, F., Toon, O. B., Pavlov, A. A., and De Sterck, H. 2005, \apj, 621, 1049.
\bibitem[Vidal-Madjar et al., (2003)]{vid03} Vidal-Madjar, A., et al. 2003, Nature, 442, 143.
\bibitem[Vidal-Madjar et al. (2004)]{vid04} Vidal-Madjar, A., et al. 2004, \apj, 604, L69.
\bibitem[Villaver and Livio (2007)]{vil07} Villaver, E., and Livio, M. 2007, \apj, 661, 1192.
\bibitem[Wood et al., (2005)]{woo05} Wood, B.E., et al. 2005, \apjs, 159, 118.
\bibitem[Yelle (2004)]{yel04} Yelle, R. 2004, Icarus, 170, 167.
\bibitem[Yelle (2006)]{yel06} Yelle, R. 2006, Icarus, 183, 508.
\bibitem[Zarka (2007)]{zar07} Zarka, P. 2007, Planet. Space Sci., 55, 598.


\end{thebibliography}
\end{document}